# Development of a dry linear compressor for GM and pulse tube cryocoolers

**J Hoehne**
Pressure Wave Systems GmbH, Haeberlstr. 8, 80337 Munich, Germany

E-mail: jens@pressurewave.de

**Abstract.** Pressure Wave Systems GmbH has developed a dry compressor for GM and pulse tube cryocoolers. The concept is based on hydraulically driven metal bellows in which the Helium working gas is compressed. The system is operated in the region of 1 kW of electrical input power and has been successfully tested with a SHI RDK-101D GM cryocooler cold head. Set-up, performance and reliability of the compressor system will be discussed.

## 1. Introduction

Conventional compressors for GM and pulse tube cryocoolers often involve oil lubricated scroll units to compress the Helium gas. Although this technology has been proven to be extremely reliable in most cryocooler systems, dry compressors might have certain advantages in terms of potential gas contamination of the cold head, maintenance and orientation constraints. Dry linear compressors have been developed for decades for Stirling and pulse tube coolers. These compressors operate at high frequency typically ranging from 20 to 100 Hz and make use of in-resonance operation. In this paper we discuss a dry low frequency linear compressor with an operating frequency of 0.5 to 2 Hz based on metal bellows technology.

## 2. Experimental

Figure 1 shows the set-up of the dry metal bellows compressor. The main components are a hydraulic pump and two metal bellows that are connected via the pump. While one bellow is compressed by the hydraulic oil on its outside the other bellow is expanded. In comparison to most set-ups in hydraulic industry this oil circuit is completely closed with the consequence that the pump needs to be operated in both directions. One of the main advantages of this arrangement is that the pump can be used as a generator when the gas pressure drives the oil.

Both bellows are equipped with check valves. In this way this set up is functionally comparable to a two-cylinder piston compressor. The main difference is that a piston compressor is usually run from hundreds to thousands of rpm where as the metal bellows compressor is run at 30 to 120 rpm. For testing purposes the electrical motor, the pump, the hydraulic cylinders as well as a heat exchangers have been water-cooled. In a commercial unit air cooling could alternatively be used.

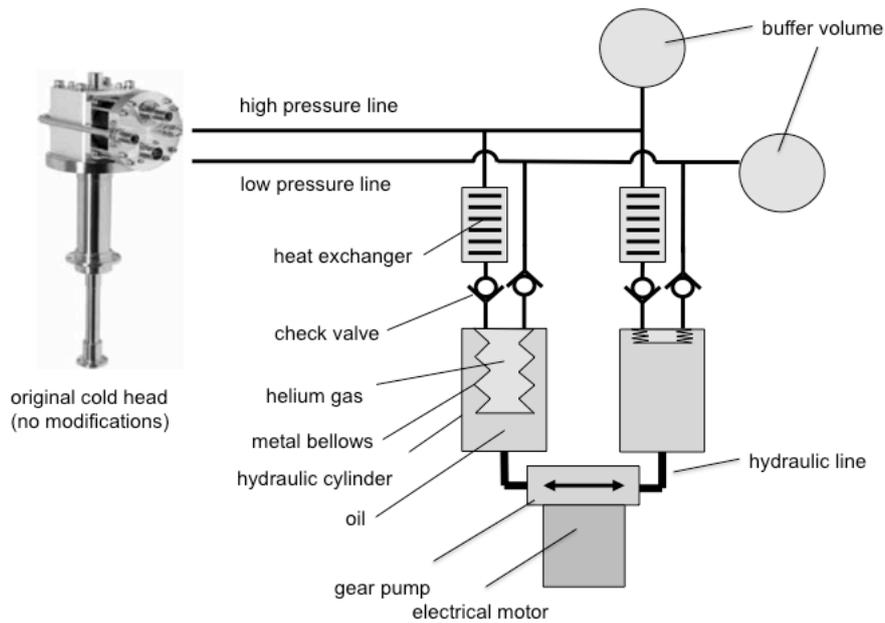

**Figure 1.** Experimental set-up of bellows compressor with cold head

## 3. Results

To test the performance of the bare compressor the system has been run with flexlines and a valve simulating the pressure drop over the cold head. Typical operating values for the Sumitomo RDK-101 cold head have been chosen for this test. Figure 2 shows the pV-performance in this set-up.

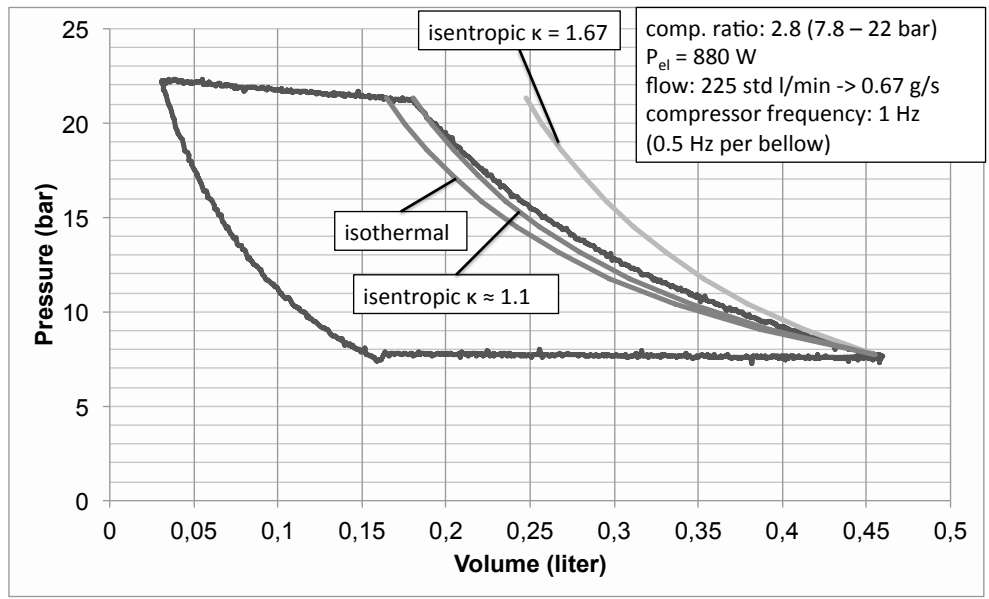

**Figure 2.** pV-performance without cold head

A typical flow of 225 standard liters per minute with a pressure drop from 7.8 to 22 bar could be realized with an electrical input power of 880 W. The compression path follows an isentropic process with a coefficient $\varkappa \approx 1.1$ which means that only small amount energy is lost due to the temperature rise of the gas.

To verify the results the compressor has been run with the SRDK-101 cold head as described in figure 1. The operating conditions were similar but not equal to the first test: the filling pressure was slightly higher resulting in a pressure range from 9.5 to 22 bar and higher Helium gas flow of 275 standard liters per minute. The electrical input power was 940 W. In this case the compression path follows an isentropic process with a coefficient of $\varkappa \approx 1.15$. Again this leads to low energetic losses during the compression process.

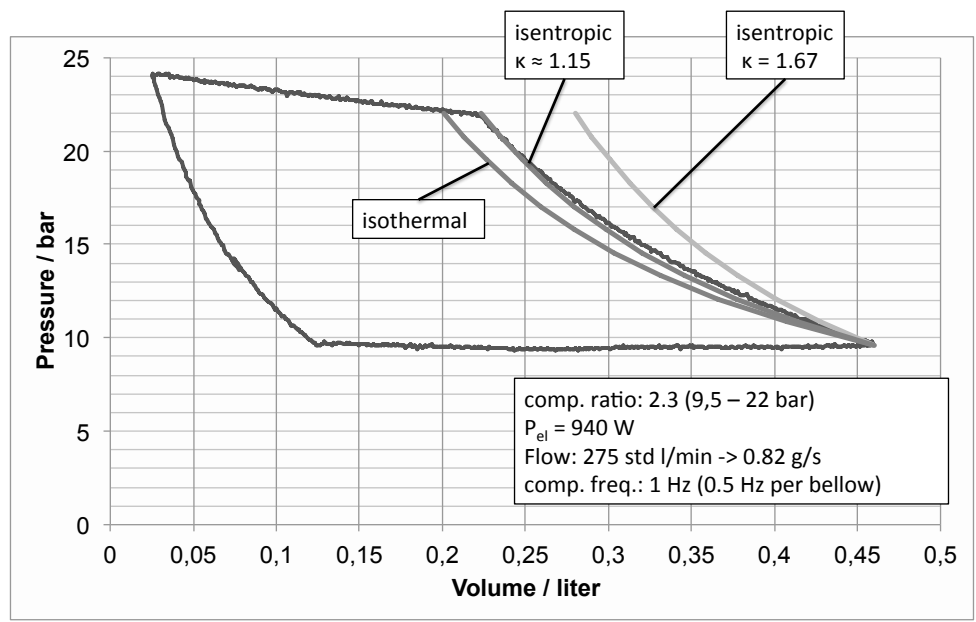

**Figure 3.** Performance with SRDK-101 cold head

The resulting cooling capacity curve of the SRDK-101 can be seen in figure 4. In the background the standard cooling capacity curve is underlayed [1]. As the cooling capacity of the SRDK-101 has been found to be pressure dependent, the capacity on the 1st stage is slightly higher than the standard curve for 1 Hz (50 Hz grid frequency) operation. The cooling capacity of the 2nd stage is similar up to 1 W and is reduced compared to the standard curve above 1W.

### 4. Discussion

*4.1. Performance and efficiency*
The bellows compressor was able to run the SRDK-101D cold head with similar performance and electrical power input as compared to the CNA-11 scroll compressor suggested by the manufacturer. The system is dry which means that no oil separation circuit has to be run and as a consequence no adsorbers have to be changed. Under normal conditions oil contamination of the cold head is impossible. For the case of a bellows failure a detection mechanism needs to be established that prevents oil leaking into the cold head and will shut down the system immediately.
The use of a bellows compressor might be even more advantageous with pulse tube cold heads as the gas quality is much more essential for safe and long-term operation.

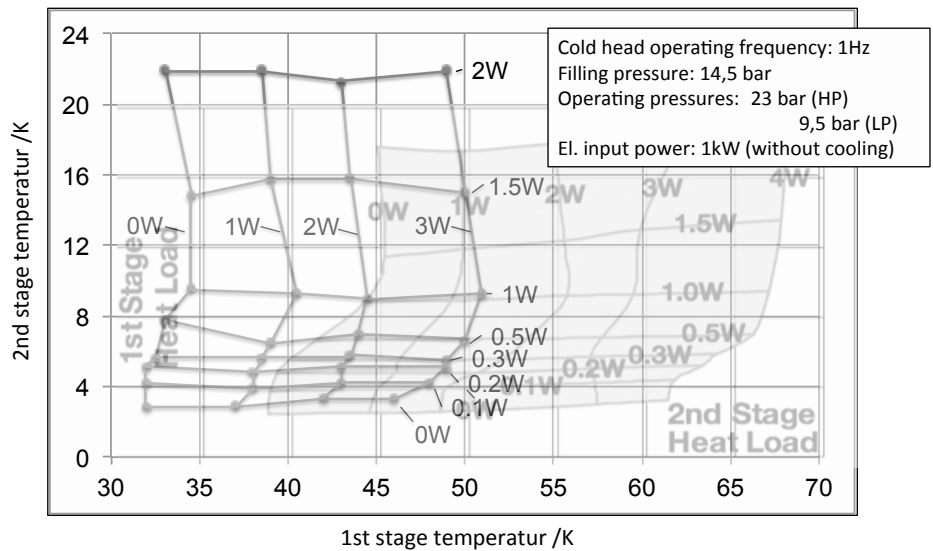

**Figure 4.** Cooling capacity of SRDK-101 cold head with compressor.

*4.2. Reliability*
It has to be shown that a bellows compressor can be as reliable as the well known and well established scroll technology. In using hydraulic technology the force on the bellows is minimized and a long lifetime can be expected. Information from bellows manufacturers lead to the assumption that operations in excess of 100 Mio cycles are well feasible without damaging the bellows structure. At a slow cycle rate of 0.5 Hz an operating time of 100 Mio cycles correspond to more than 55.000 hours or more than 6 years of continuous running time.
Commercial check valves have been used for first tests but turned out to fail eventually. We have started our own check valve development and have verified their operation to 40 Mio cycles – assessment ongoing.

**5. Summary and outlook**
A new type of dry compressor based on hydraulics and metal bellows technology has been developed. It may have future use in driving GM and pulse tube cryocoolers in applications where dry and/or low vibration operation is desired or the compressor needs to operate orientation independent. Its reliability and its advantages to conventional scroll based compressor have yet to be determined. The concept allows for straight forward upscaling of the electrical power from now 1 kW to 8-12 kW, which would be sufficient to drive the largest cold heads in todays marketplace.

**6. Acknowledgement**
The author would like to thank Sumitomo (SHI) Cryogenics of Europe GmbH and Cryophysics GmbH, Germany as well as Sumitomo (SHI) Cryogenics of Europe Ltd., UK, for their support.
The discussions with M. Bühler and the support by Low Temperature Solutions UG, Ismaning, have been very valuable for this work.
The speed and accuracy in making mechanical parts by G. Töpelt is greatly appreciated.
Thanks to Nick N. who shared even more music of a linear compressor in his office.
This work was happily supported by WMH, FH and JH.

**7. References**
[1]   Sumitomo Heavy Industries, Ltd., Cryocooler Product Catalogue, 05/12